\documentclass[epsf,usegraphicx]{mn2e}

\usepackage{xspace}

\title[Followup observations of 4 newly discovered short period variables]
{RApid Temporal Survey
- RATS II: Followup observations of 4 newly discovered short period variables}

\author[Ramsay, Napiwotzki, Hakala \& Lehto]
{Gavin Ramsay$^{1}$, 
Ralf Napiwotzki$^{2}$, Pasi Hakala$^{3}$,  Harry Lehto$^{3,4}$\\
$^{1}$Mullard Space Science Lab, University College London,
Holmbury St. Mary, Dorking, Surrey, RH5 6NT, UK\\
$^{2}$Centre of Astrophysics Research, STRI, University of Hertfordshire,
Hatfield, AL10 9AB, UK\\
$^{3}$Tuorla Observatory, University of Turku, V\"ais\"al\"antie 20, FIN-21500
Piikki\"o, Finland\\
$^{4}$NORDITA, Blegdamsvej 17, DK-2100. Copenhagen, Denmark
}

\date{Accepted 2006 June 23. Received 2006 June 22; in original form 2006 May 17}

\begin{document}
\outer\def\gtae {$\buildrel {\lower3pt\hbox{$>$}} \over 
{\lower2pt\hbox{$\sim$}} $}
\outer\def\ltae {$\buildrel {\lower3pt\hbox{$<$}} \over 
{\lower2pt\hbox{$\sim$}} $}
\newcommand{\ergscm} {ergs s$^{-1}$ cm$^{-2}$}
\newcommand{\ergss} {ergs s$^{-1}$}
\newcommand{\ergsd} {ergs s$^{-1}$ $d^{2}_{100}$}
\newcommand{\pcmsq} {cm$^{-2}$}
\newcommand{\ros} {\sl ROSAT}
\newcommand{\exo} {\sl EXOSAT}
\def\rchi{{${\chi}_{\nu}^{2}$}}
\newcommand{\Msun} {$M_{\odot}$}
\newcommand{\Mwd} {$M_{wd}$}
\def\Mdot{\hbox{$\dot M$}}
\def\mdot{\hbox{$\dot m$}}
\newcommand{\teff}{\ensuremath{T_{\mathrm{eff}}}\xspace}
\newcommand{\ratus}{RAT\,J0455+1305\xspace}

\maketitle

\begin{abstract}

The RApid Temporal Survey (RATS) is a survey to detect objects whose
optical intensity varies on timescales of less than $\sim$70 min.  In
our pilot dataset taken with the INT and the Wide Field Camera in Nov
2003 we discovered nearly 50 new variable objects. Many of these
varied on timescales much longer than 1 hr. However, only 4 objects
showed a modulation on a timescale of 1 hour or less. This paper
presents followup optical photometry and spectroscopy of these 4
objects. We find that RAT J0455+1305 is a pulsating (on a period of
374 sec) subdwarf B (sdB) star of the EC\,14026 type. We have modelled
its spectrum and determine $\teff = 29,200\pm 1900$\,K and $\log g =
5.2 \pm 0.3$ which locates it on the cool edge of the EC\,14026
instability strip. It has a modulation amplitude which is one of the
highest of any known EC 14026 star. Based on their spectra,
photometric variability and their infra-red colours, we find that RAT
J0449+1756, RAT J0455+1254 and RAT J0807+1510 are likely to be SX Phe
stars - dwarf $\delta$ Sct stars. Our results show that our observing
strategy is a good method for finding rare pulsating stars.

\end{abstract}


\section{Introduction}

\begin{figure*}
\begin{center}
\setlength{\unitlength}{1cm}
\begin{picture}(8,11)
\put(-1.3,-0.4){\includegraphics{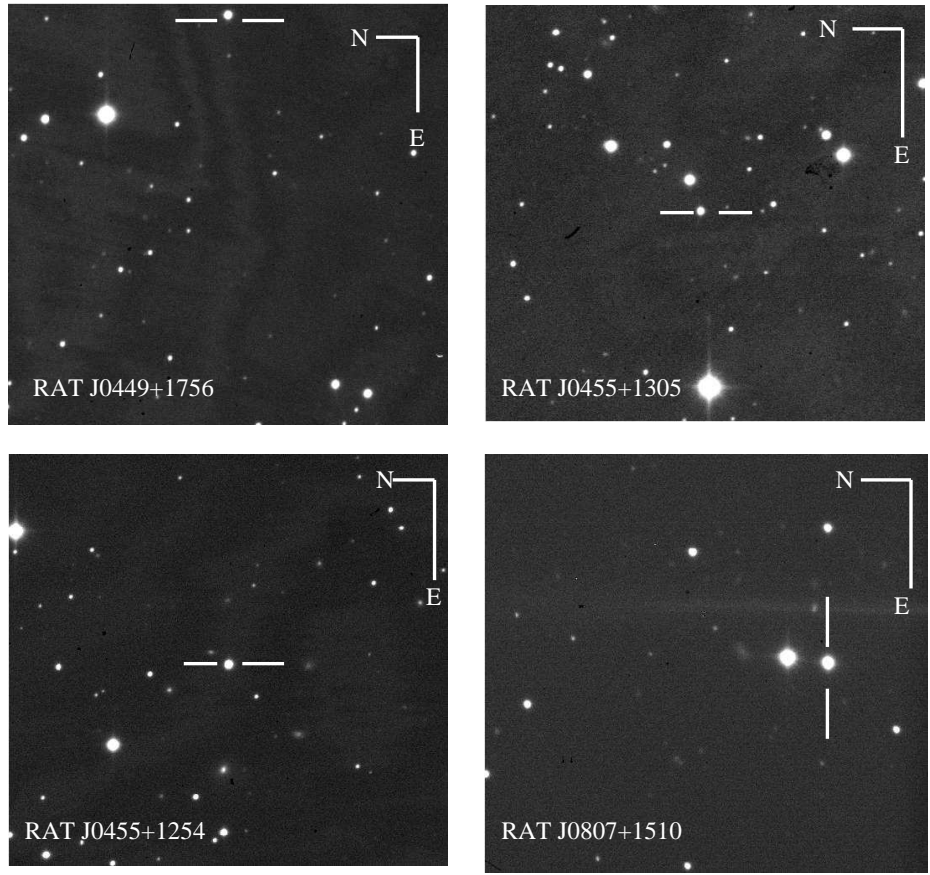}}
\end{picture}
\end{center}
\caption{The finding chart for each of the sources presented in this 
paper. The location of each source is centered between the thick lines.
The size of the field is 3$^{'}\times3^{'}$ and was taken in white light
using the INT.}
\label{finding}
\end{figure*}

The aim of the RApid Temporal Survey (RATS) is to discover objects
whose optical intensity varies on timescales of a few mins to several
hours (Ramsay \& Hakala 2005). The prime aim is to discover
interacting ultra-compact binary systems - systems consisting of two
degenerate (or semi-degenerate) stars orbiting around a common center
of gravity - with binary orbital periods less than $\sim$70 mins. Our
pilot set of data was obtained from La Palma in Nov 2003 using the
Isaac Newton Telescope and the Wide Field Camera: it covered 3 square
degrees and reached a depth of $V\sim$22.5. Nearly 50 sources were
found to show significant intensity variations: none were previously
known variable objects (Ramsay \& Hakala 2005). However, only 4
objects showed modulations which varied on periods of approximately 1
hour or less. Table \ref{sources} shows the positions and modulation
periods, while in Figure \ref{finding} we show the finding chart for
each object. This paper reports followup observations of these 4
systems, the aim being to determine their nature.

\section{Observations}

We obtained photometric and spectroscopic observations of our targets
using the Nordic Optical Telescope (NOT) in visitor mode and the
William Herschel Telescope (WHT) in service mode: both are located on
the island of La Palma. Photometric observations were also obtained
using the Greek Kryoneri 1.2m telescope - the data was of lower signal
to noise than the data obtained using the NOT and so are therefore not
discussed in detail. A log of our observations is shown in Table
\ref{log}.  The data taken using the NOT were made using ALFOSC which
allows both photometric and spectroscopic data to be taken. The CCD
was windowed to reduce readout time.

Photometric data taken using the NOT were made using white light (in
the case of RAT J0455+1305) and in $B,R$ filters (RAT J0449+1756).
The exposure times were 15 sec, 20 sec and 30 sec for the white light,
$R$ band and $B$ band data respectively. In all bands the dead time
was $\sim$8 sec.

Spectra were obtained of all 4 targets and taken using relatively low
resolution grisms (Table \ref{log}). This resulted in spectra with
resolutions between $\sim$5--18\AA. The spectra obtained using the WHT
were made using both the red and blue arms of ISIS.  All data were
reduced and analysed using the usual procedures.

\begin{table}
\caption{The objects for which we present photometric and spectroscopic data. 
We show the 
co-ordinates (J2000), the modulation period and the $V$ or $R$ band mag.}
\begin{center}
\begin{tabular}{l@{\hspace{5pt}}c@{\hspace{2pt}}cr@{}l@{\hspace{3pt}}r}
\hline
Source &  RA & Dec & \multicolumn{2}{c}{Period} & Mag\\
       &     &     & \multicolumn{2}{c}{(min)} &     \\
\hline
RAT J0449+1756 & 04 49 52.3 &  +17 56 37.8 & 40 &  & $R$=16.1\\
RAT J0455+1305 & 04 55 15.2 &  +13 05 29.7 & 6  & .23 & $V$=17.2\\
RAT J0455+1254 & 04 55 16.5 &  +12 54 10.5 & 66 &  & $V$=16.0\\
RAT J0807+1510 & 08 07 00.5 &  +15 10 58.2 & 60 &  & $V$=15.4\\
\hline
\end{tabular}
\end{center}
\label{sources}
\end{table}

\begin{table}
\caption{The observation log. The NOT refers to the Nordic Optical
Telescope, the WHT the William Herschel Telescope and Kry the Kryoneri
Telescope.}
\begin{center}
\begin{tabular}{lccrr}
\hline
Source & Date  & Telescope & Band & Time \\
\hline
{\it Photometry} & & & & \\
RAT J0449+1756 & 2004 12 05 & NOT & $B,R$ & 6.1 hr\\ 
               & 2004 12 06 & NOT & $B$   & 1.4 hr\\
               & 2004 12 08 & NOT & $B$   & 1.5 hr\\
RAT J0455+1305 & 2004 02 17 & NOT & $WL$  & 1.2 hr\\
               & 2004 12 07 & NOT & $WL$  & 3.0 hr\\
               & 2004 12 08 & NOT & $WL$  & 1.7 hr\\
               & 2005 12 08 & Kry & $WL$    & 4.5 hr\\
\hline
{\it Spectroscopy} & & & & \\
RAT J0449+1756 & 2004 09 24 & WHT & R300B & 400 sec\\
               &            &     & R316R & 400 sec\\
RAT J0455+1305 & 2004 12 07 & NOT & 4/300 & 600 sec\\
RAT J0455+1254 & 2004 09 06 & NOT & 8/600 & 900 sec\\
               &            &     & 4/300 & 300 sec\\
RAT J0807+1510 & 2004 01 20 & NOT & 11/200 & 300 sec \\
\hline
\end{tabular}
\end{center}
\label{log}
\end{table}

\section{Results}

\subsection{Photometric Results}

\begin{figure}
\begin{center}
\setlength{\unitlength}{1cm}
\begin{picture}(8,7)
\put(-1.4,-0.3){\includegraphics{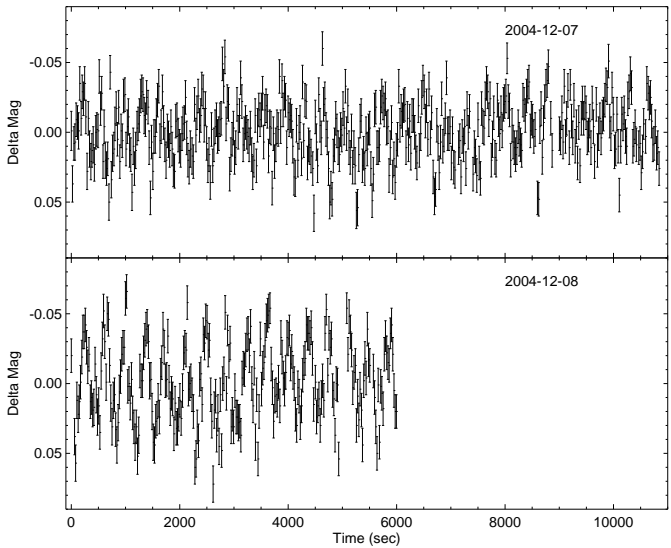}}
\end{picture}
\end{center}
\caption{Photometric observations of RAT J0455+1305 made in white
light using the NOT. The exposure time was 15 sec and the deadtime was
$\sim$8 sec.}
\label{rat0455_1305_lc}
\end{figure}

\begin{figure}
\begin{center}
\setlength{\unitlength}{1cm}
\begin{picture}(8,5.4)
\put(-0.8,-0.4){\includegraphics{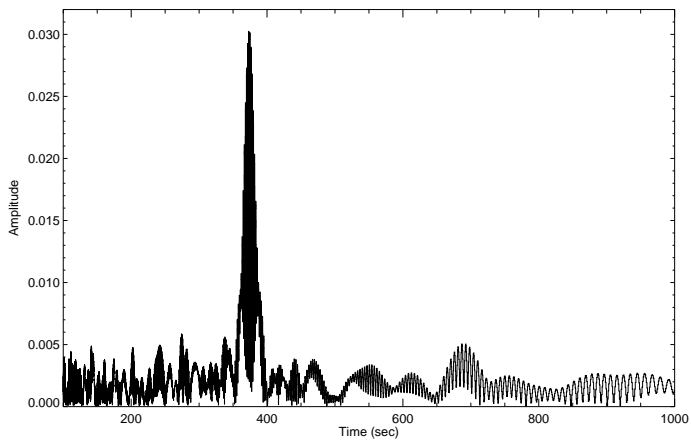}}
\end{picture}
\end{center}
\caption{The amplitude spectrum of the light curve of RAT J0455+1305
taken on 7 \& 8 Dec 2004.}
\label{rat0455_1305_power}
\end{figure}

RAT J0455+1305 showed prominent intensity variations with an amplitude
of 50 mmag in white light on a timescale of 374 sec (Ramsay \& Hakala
2005). We show the light curve taken in Dec 2004 from the NOT in
Figure \ref{rat0455_1305_lc} and we show its amplitude spectrum in
Figure \ref{rat0455_1305_power}. This new data again shows a prominent
period close to 374 sec, with the strongest candidate periods being
373.1 and 374.8 sec. 

\begin{figure}
\begin{center}
\setlength{\unitlength}{1cm}
\begin{picture}(8,11)
\put(-0.5,-1.7){\includegraphics{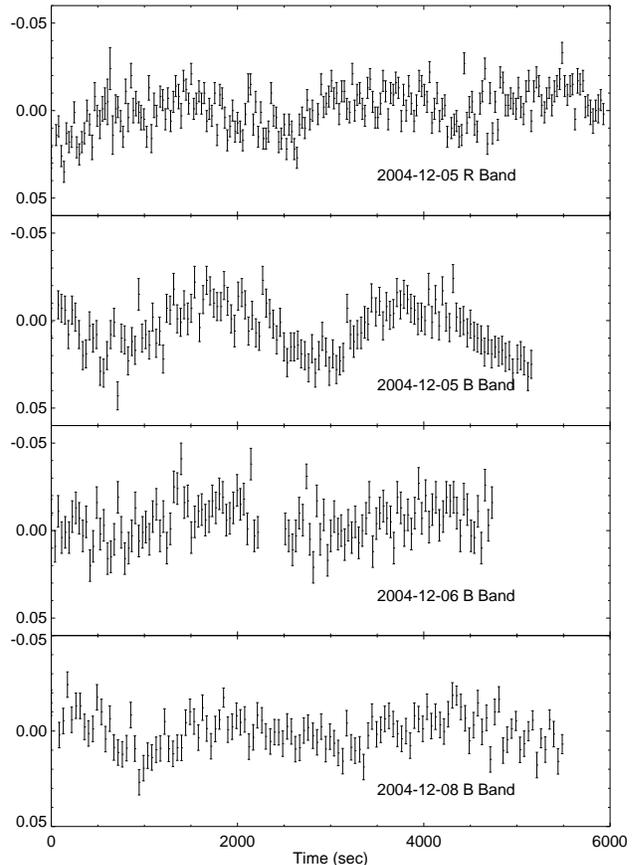}}
\end{picture}
\end{center}
\caption{Photometric observations of RAT J0449+1756 made using the
NOT. The date of the observation and the filter are indicated in each panel.}
\label{rat0449}
\end{figure}

Ramsay \& Hakala (2005) showed that RAT J0449+1756 varied in white
light on a period of either $\sim$40 or 80 min in a sinusoidal
manner. We obtained a short observation in the $R$ band and three
observations in the $B$ band (Figure \ref{rat0449}). These new
observations show a variation in both $R$ and $B$ bands. As found in
the original INT data, the modulation period is close to 40 min (or
twice this period) with the shape of the $B$ band modulation period
and amplitude varying from 0.03--0.05 mag.

The light curves of both RAT J0455+1254 and RAT J0807+1510 show a
prominent modulation on a period just over 1 hr. Neither shows
strictly sinusoidal light curves, rather they show narrower peaks and
broad minima (Ramsay \& Hakala 2005).

\subsection{Spectroscopic Results}

\begin{figure}
\begin{center}
\setlength{\unitlength}{1cm}
\begin{picture}(8,12)
\put(-0.5,-0.8){\includegraphics{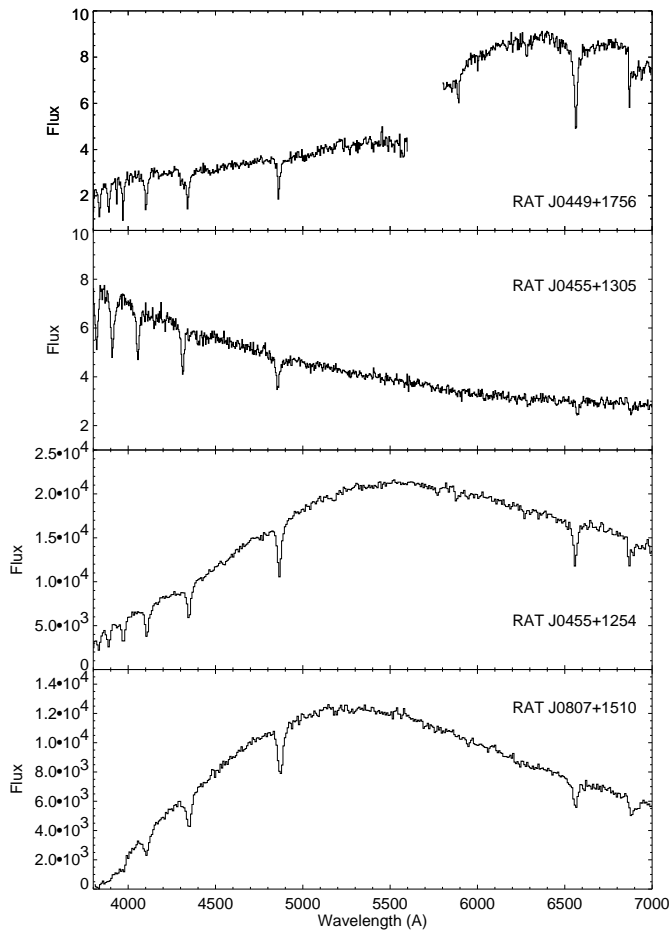}}
\end{picture}
\end{center}
\caption{The optical spectra of the sources discussed in this paper. 
Apart from RAT J0449+1756 which was obtained using the WHT, they were
obtained using the NOT.}
\label{spectra}
\end{figure}

Single epoch spectra were obtained of each source using low resolution
gratings (cf Table \ref{log}). Apart from RAT J0455+1305 the spectra
were not flux calibrated and therefore partly reflect the instrumental
response. The spectra are shown in Figure \ref{spectra}. The first
point to note is that none of the sources show emission lines. Apart
from RAT J0455+1305 (which shows a very blue spectrum) all the sources
show spectra typical of early F-type main sequence stars.

Compared to B-type main sequence stars RAT J0455+1305 shows relatively
broad hydrogen absorption lines. Moreover, the presence of hydrogen
rules out a PG 1159 star identification (He-rich hot pre-white dwarf
stars), while the lines are too narrow to be a white dwarf. It is
unlikely that it is a cool white dwarf where Stark broadening
disappears.  Its spectrum together with the prominent optical
pulsations strongly imply that it is an EC 14026 star. Around 20 of
these subdwarf B stars are known: see O'Donoghue et al (1998) for a
review.  In the next section we perform a more detailed analysis of
these spectra.

\section{The nature of these objects}

The fact that none of the systems reported in this paper show emission
lines suggests that accretion was not occurring in these objects at
the epoch of our observations. If any of these systems are stellar
binary systems the secondary is therefore not filling its Roche
lobe. Further, the fact that all objects show hydrogen rather than
helium lines rules out the systems being ultra-compact binary
systems. We have searched X-ray, UV and radio catalogues and do not
find objects in the same location as our sources. Our sources can be
split up into one system (RAT\,J0455+1305) which shows short period
pulsations (374\,sec) and a spectrum typical for a hot B-type star,
and three F-type stars with longer period light variations. We now
discuss the nature of these systems.

\subsection{RAT J0455+1305}

Our photometric and spectroscopic data strongly suggests that RAT
J0455+1305 is an EC 14026 star. They are non-radial pulsating subdwarf
B (sdB) stars and are named after the prototype EC\,14026$-$2647. SdB
stars in general are identified with stars on the extreme horizontal
branch (EHB; Heber 1986). They are core helium burning stars, which
have lost almost their complete hydrogen envelope at the end of the
first red giant branch. Therefore hydrogen shell burning cannot be
sustained and after the end of their core helium burning phase these
stars are not able to ascend the asymptotic giant branch, but evolve
directly to the white dwarf cooling sequence.

Variable EC\,14026 stars are a subset of the known sdB stars found in
a specific region of the Hertzsprung-Russell diagram (Charpinet et
al.\ 2001). Usually a number of periods ranging from 90\,sec up to
600\,sec are observed, although in some cases only one dominant period
is detected with other periods being much weaker (eg HS\,0702+6043
mentioned below; Dreizler et al. 2002). However, we could only
identify one clear pulsation period in our photometric data. This
could be explained by our relatively short dataset. It is likely that
observations with medium sized telescopes covering longer time spans
will reveal additional frequencies.

\begin{figure}
\begin{center}
\setlength{\unitlength}{1cm}
\begin{picture}(8,10.8)
\put(-1.,-1.1){\includegraphics{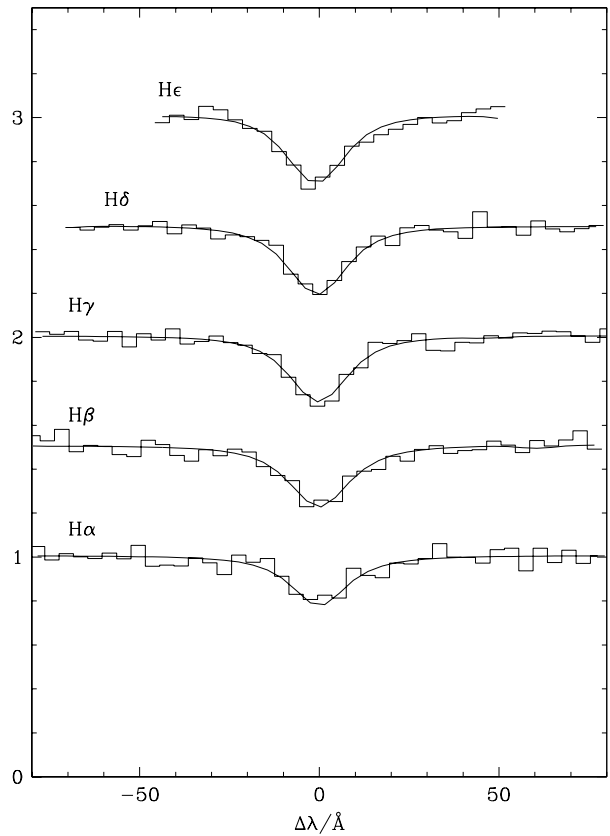}}
\end{picture}
\end{center}
\caption{We show the Balmer lines of \ratus overlaid with the best model 
fits using {\tt FITSB2} (Napiwotzki et al. 2004). For each spectral line the 
continuum has been normalised to 1.0 and each spectrum after H$_{\alpha}$
has been shifted up by 0.5 units.}
\label{balmerfit}
\end{figure}

\begin{figure}
\begin{center}
\setlength{\unitlength}{1cm}
\begin{picture}(8,6)
\put(-0.6,-0.5){\includegraphics{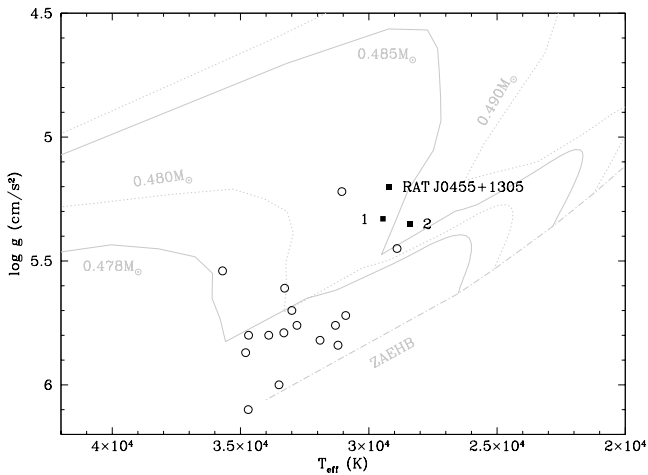}}
\end{picture}
\end{center}
\caption{Evolutionary tracks for moderately metal poor EHB stars
($\mathrm{[Fe/H]}=-$0.47) from Dorman et al (1993). The tracks are
labeled with the stellar masses. The zero-age EHB (ZAEHB) is indicated
by the dash-dotted line (extrapolated to hotter temperatures for
illustrative purposes assuming constant luminosity).  The position of
EC\,14026 stars from the compilation of Charpinet et al.\ (2001) are
plotted as open circles. RAT\,J0455+1305 and the very similar objects
HS\,0702+6043 (1) and Balloon\,090100001 (2) are indicated by the
filled squares.}
\label{tracks}
\end{figure}

We modelled the spectrum of \ratus (Figure \ref{spectra}) using a
model atmosphere analysis with a grid of hydrogen and helium composed
NLTE model atmospheres (Napiwotzki 1997). Model profiles were fitted
to the Balmer lines to derive the effective temperature \teff, surface
gravity $\log g$ and metallicity. The signal to noise of the spectrum
is not high enough to allow a determination of the helium
abundance. Therefore, we adopted a value of
$n_{\mathrm{He}}/n_{\mathrm{H}} = 3\times 10^{-3}$ which is typical
for sdB stars with the parameters of \ratus (Edelmann et al.\
2003). However, our fit result is not sensitive to this adopted
abundance. The Balmer lines were fitted using {\tt FITSB2} (Napiwotzki
et al.\ 2004). The error limits were determined with a bootstrapping
method. Our best fit is shown in Figure \ref{balmerfit} and gives
$\teff = 29200\pm 1900$\,K and $\log g = 5.2 \pm 0.3$. We compare the
position of \ratus in a \teff-$\log g$ diagram with evolutionary
tracks of extreme horizontal branch stars in Figure
\ref{tracks}. \ratus has already left the horizontal branch and will
now evolve to higher temperatures and become a white dwarf.  The
parameters of \ratus places it at the cool side of the so-called
EC\,14026 instability strip (Charpinet et al.\ 2001). The longest
pulsation periods are seen in the coolest EC\,14026 stars, which fits
well with the period of 374 sec observed in \ratus.

Most EC\,10426 stars show only intensity variations of a few mmag
(cf.\ compilation in Table~1 of Charpinet et al.\ 2001). Our observed
amplitude of 50\,mmag makes \ratus one of the highest amplitude
EC\,14026 stars. Two EC\,10426 stars with very similar properties are
currently known: HS\,0702+6043 (Dreizler et al.\ 2002) and
Balloon\,090100001 (Oreiro et al.\ 2004). These are shown as objects 1
and 2 in Figure \ref{tracks}. Their dominant frequencies are 363\,sec
and 356\,sec, respectively, which are interpreted as non-radial p-mode
pulsations. The corresponding amplitudes of 29\,mmag and 53\,mmag were
(until now) the highest observed in EC\,10426 stars. One exciting
aspect of these two objects is that they are the only known hybrid
sdBV pulsators showing long period ($\approx 1$\,hr) g-modes as well
(Schuh et al. 2006; Baran et al.\ 2005).  The very similar parameters
we derive for RAT\,J0455+1305 make it a very good candidate for being
the third object in this class of hybrid pulsators. Interesting
results can be expected from dedicated photometric observations.

\subsection{RAT J0449+1756, RAT J0455+1254 \& RAT J0807+1510}

The spectra of RAT J0449+1756, RAT J0455+1254 \& RAT J0807+1510 are
consistent with early F-type main sequence stars. Their photometric
light curves are very similar to SX\,Phe type variable stars (dwarf
$\delta$ Cephei stars). SX Phe stars are related to the $\delta$ Scuti
pulsators showing modulation periods in the range of less than one
hour to several hours and a modulation amplitude between 0.01 mag and
$\sim$1 mag. The photometric data of RAT J0449+1756 showed a
modulation period of 40 min, although it is possible the true period
was twice this if the light curve was intrinsically double peaked. For
a SX Phe interpretation, the period of RAT J0449+1756 would therefore
be 40 min.

\begin{figure}
\begin{center}
\setlength{\unitlength}{1cm}
\begin{picture}(8,6)
\put(-1.5,-0.5){\includegraphics{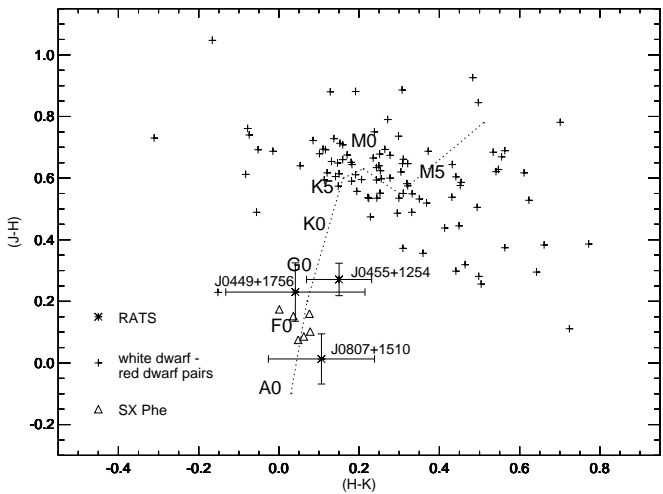}}
\end{picture}
\end{center}
\caption{The infra-red colours of the 3 of our targets (asterix's),
red dwarf - white dwarf binaries (taken from the source list of
Wachter et al 2003) and SX Phe field stars (the infrared colours are
taken from the 2MASS archive).}
\label{2mass_ms}
\end{figure}

To test if the colours of our sources are consistent with known SX Phe
stars we obtained their infra-red colours from the
2MASS\footnote{http:www.ipac.caltech.edu/2mass} catalogue (Table
\ref{2mass}). We show in Figure \ref{2mass_ms} the main sequence in
the $(J-H), (H-K)$ colour plane (taken from Wachter et al 2003). RAT
J0449+1756, RAT J0455+1254 and RAT J0807+1510 fall close to the
expected position for early F-type stars. We also show the location of
known SX Phe stars in this colour plane: they occupy a very similar
location to our newly discovered sources. For comparison we also show
in Figure \ref{2mass_ms} the colours of white dwarf - red dwarf
binaries (these objects were taken from the list of Raymond et al 2003
and their infrared colours obtained from the 2MASS catalogue).  These
binaries are clearly separated from our targets suggesting that they
do not have low mass red dwarf companions.

We modelled the spectra of our F-type stars using a similar approach
to that undertaken for \ratus.  It was not possible to constrain the
mass of these systems so gravity was fixed at $\log g = 4.0$, which is
a typical value for the sample of SX\,Phe and large amplitude $\delta$
Scuti stars investigated by McNamara (1997).  A grid of LTE models was
calculated with the {\tt ATLAS\,9} code (Kurucz 1992) with convective
overshooting switched off. Spectra were subsequently calculated with
the {\tt LINFOR} line-formation code (Lemke 1991). Data for atomic and
molecular transitions from the Kurucz line list was taken into
account. For the actual fitting we used again the {\tt FITSB2} routine
(Napiwotzki et al.\ 2004).

\begin{figure}
\begin{center}
\setlength{\unitlength}{1cm}
\begin{picture}(8,6.7)
\put(-1.0,-0.5){\includegraphics{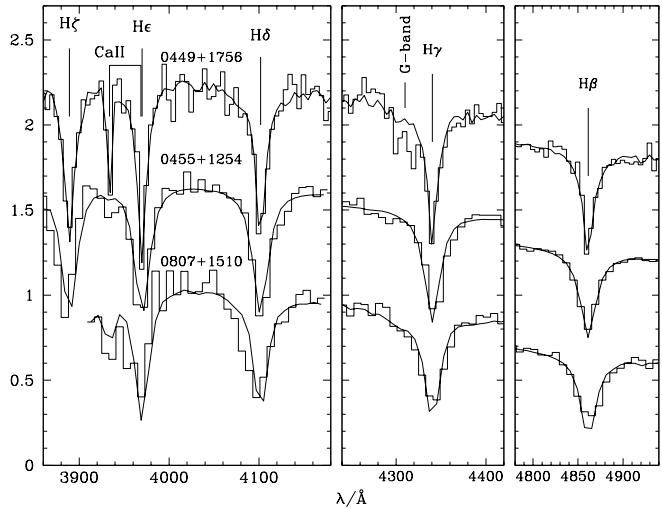}}
\end{picture}
\end{center}
\caption{The spectral fits to RAT J0449+1756, RAT J0455+1254 and 
RAT J0807+1510. The derived temperatures and masses are shown in Table 
\ref{sx_spec}.}
\label{fstar_fits}
\end{figure}

The Balmer lines are good temperature indicators for F stars.  Two
features in our spectra (Figure \ref{fstar_fits}) can be used as
metallicity indicators: the Ca H+K lines and the G-band in the blue
wing of H$\gamma$ (mostly CH absorption). Our best fits are shown in
Figure \ref{fstar_fits} and the results are given in Table
\ref{sx_spec}. Although metallicities are not strongly constrained,
all 3 stars appear to have metallicities below solar. The strongest
indication for this is found in RAT\,J0455+1254 with very weak Ca\,H+K
and G-band lines.  A good fit to the observed G-band of RAT\,0449+1756
would require higher metallicity than indicated by the Ca lines. The
short wavelength part of the spectrum of RAT\,0807+510 has low signal
to noise so that the metallicity determination relies on the G-band.

We note that our fits aim to provide an estimate of stellar
temperature and metallicity to facilitate the classification. They are
not meant to substitute a comprehensive spectral analysis.  The
derived temperatures are within the range found by McNamara (1997) for
SX\,Phe and high amplitude $\delta$ Scuti stars. However, a closer
inspection reveals that our temperatures (Table \ref{sx_spec}) are
slightly cooler then predicted for the measured periods (Figure ~2
and~3 in McNamara): 7200-7400\,K vs.\ 7600-8000\,K. Given the limited
quality of our spectra, we do not think that this is a significant
discrepancy. Better observations would be required for a more accurate
parameter determination.

\begin{table}
\caption{The infrared colours of our targets. Data taken
from the 2MASS web site (RAT J0455+1305 is not in the 2MASS database). The 
error in parenthesis is the error on the last 3 digits.}
\begin{center}
\begin{tabular}{lrrr}
\hline
Source &  $J$ & $H$ & $K$ \\
\hline
RAT J0449+1756 & 15.405(051) & 15.175(080) & 15.134(154) \\
RAT J0455+1254 & 14.563(033) & 14.292(041) & 14.142(070) \\
RAT J0807+1510 & 14.817(038) & 14.804(072) & 14.698(111) \\
\hline
\end{tabular}
\end{center}
\label{2mass}
\end{table}

\begin{table}
\caption{The temperature and metallicity for the 3 sources showing 
F-type stellar spectra derived
from model fits to their optical spectra.}
\begin{center}
\begin{tabular}{lccrr}
\hline
Source &  $T_{eff}$ &  $\log$ [M/H]\\
\hline
RAT J0449+1756 & 7270K & -0.7\\
RAT J0455+1254 & 7460K & -3.5\\
RAT J0807+1510 & 7410K & -0.6\\
\hline
\end{tabular}
\end{center}
\label{sx_spec}
\end{table}

We estimate their approximate distance using the relationship of Nemec
et al (1994): $M_{V}=0.36-2.56 \log P +0.32\mathrm{[Fe/H]}$, which is
appropriate for SX Phe stars pulsating in the fundamental mode. For a
period of 1 hr and $\mathrm{[Fe/H]}$=--2, the above relation predicts
$M_{V}$=3.3. For the apparent brightness of RAT J0455+1254 and RAT
J0807+1510 (Table \ref{sources}) we find approximate distance of 3.5
and 2.6 kpc respectively. For Galactic latitudes of $-18^{\circ}$ and
+23$^{\circ}$ we find that the sources are displaced from the Galactic
plane by 1.1kpc and 1.0kpc for RAT J0455+1254 and RAT J0807+1510
respectively: this is several scale-heights above the Galactic plane.

\section{Conclusions}

We have obtained followup optical photometry and spectroscopy of 4
sources which were found to show intensity modulations on timescales
of less than $\sim$1 hr in our pilot RATS dataset. We find that the
source which shows the shortest period (374 sec), RAT J0455+1305, is
an EC 14026 star.  We have modelled its Balmer absorption lines and
find a temperature of 29200K$\pm$1900 K and a gravity of $\log
g=5.2\pm0.3$. This places RAT J0455+1305 at the cool side of the EC
14026 instability strip. The other sources have spectra which indicate
temperatures between 7200-7500K. This together with their optical
light curves and infra-red colours strongly suggest they are SX Phe
stars.

There are currently around 20 EC 14026 stars, while there are less
than 20 currently known SX Phe stars which are not associated with
globular clusters, open clusters or external galaxies. The usual
method of discovering EC 14026 stars is a three-step
process. Candidate stars are discovered using colour or prism
surveys. Spectroscopy is then required to confirm their classification
and determine their temperature and gravity. Lastly, high speed
photometry is needed to search for variability. We have found that our
`reverse' strategy is a good method of discovering variable EC 14026
stars.  Further, our strategy appears to be a good method of
discovering SX Phe stars, which should help determine their space
density and period distribution.

\section{Acknowledgements}

Observations were made using the Nordic Optical Telescope and the
William Herschel Telescope on La Palma and the Kryoneri Telescope in
Greece.  We gratefully acknowledge the support of each of the
observatories staff. Some of the data presented here have been taken
using ALFOSC, which is owned by the Instituto de Astrofisica de
Andalucia (IAA) and operated at the Nordic Optical Telescope under
agreement between IAA and the NBIfAFG of the Astronomical Observatory
of Copenhagen. We thank Darragh O'Donoghue, Uli Heber and Don Pollacco
for useful comments on EC 14026 stars.  R.N.\ acknowledges support by
a PPARC Advanced Fellowship.

{}

\end{document}